# A Combinatorial Auction Design for Formulary Positions

by

Lawrence W. Abrams, Ph.D.

April 2, 2025

## Summary

The Wyden-Grassley 2019 Senate Staff Report confirms that rebate contracts between pharmaceutical companies (Pharma) and the Big 3 pharmacy benefit managers (PBMs) contain a combinatorial bid-menu for formulary positions featuring both exclusive and shared positions. Pharma bid-menus often include a separate bid option for "an incremental base rate" if the PBM outright excludes a named competing drug. The standard bid basis is a percentage off a publicly available list price.

The purpose of this paper is to apply the economics field of market design to develop a simple algebraic and graphic model of a combinatorial auction for formulary position assignments.

This paper is evidence of economist Ran Spiegler's observation in his book *The Curious Culture of Economic Theory* that market design economics blurs the lines between an economist and the market designer. In places, this economist becomes a market designer suggesting the following ways PBMs can improve their auction design:

1. a test for anticompetitive incremental exclusionary rebate bids.
2. a more incentive-compatible auction by limiting shared position bid-downs.
3. banning lump sum, bundled, market share, and all other non-linear rebate bid bases.

## Disclosures:

I have not received any remuneration for this paper nor have I financial interest in any company cited in this working paper. I have a Ph.D. in Economics from Washington University in St. Louis and a B.A. in Economics from Amherst College. Other papers on PBMs can be found on my website  https://labrams.com

**I. Introduction**

The purpose of this paper is to apply the economics field of market design to the exchange of rebates for favored formulary positions. We develop a simple algebraic and graphic model of a combinatorial position auction for formulary assignments. The model includes a bid option for outright exclusion of named competing drugs. Based on Aghion and Bolton's (AB) foundational paper on vertical contracting with incomplete information, we view this option as liquidated damages.[1]

The market designers here are the Big 3 pharmacy benefit managers (PBMs) -- CVS Caremark, Aetna Express Scripts, and United Healthcare OptumRx. They are countervailing intermediaries managing prescription (Rx) drug benefit plans as agents on behalf of plan sponsors.

There are places in the paper where we transition from what is going on to what we think should be going on based on economics of good auction design. Unlike other economic models with passive Walrasian market makers, our market design model has PBMs as auction designers responsible for choices relating to bid basis, bid menus, and the winners' determination equation. PBMs are capable of determining if liquidated damage bids are anticompetitive as they possess the required data for this test.

---
[1] Aghion, P., and P. Bolton. 1987. "Contracts as a Barrier to Entry." American Economic Review, 77(3): 388–401.
https://business.columbia.edu/sites/default/files-efs/pubfiles/2018/contracts%20a%20barrier%20too%20

This paper is evidence of economist Ran Spiegler's observation in his book *The Curious Culture of Economic Theory* that the lines between economist and the market designer can become blurred.[2]   In Chapter 8, Spiegler takes the reader through his oscillation between an equilibrium model and a market design model in analyzing Google's ad position auction.  Spiegler compares his experience with an equilibrium model to that of a market design approach

> "...which regards the economist as an "engineer" that designs details of the market institution." [3]
> and "…Impels the analyst to expand the set of the instruments at the designer's disposal…" [4]

This paper is similar to Spiegler's experience. In places, this economist becomes a market designer suggesting ways PBMs can improve their auction design by:

1. developing a test for anticompetitive incremental exclusionary rebate bids.
2. designing a more incentive-compatible auction by limiting incumbent drug shared position bid-downs.
3. explicitly banning lump sum, bundled, market share, and all other non-linear rebate bid bases.

---

2 RAN SPIEGLER, The Curious Culture of Economic Theory, The MIT Press, April 2024.
3 SPIEGLER, *supra*, p 157
4 *ibid.*

II. Conceptualization

We had an "aha" moment in 2023 when we read sections of The Grassley-Wyden Senate Staff Report (hereafter, GW Report) based on over 1,700 pages of internal documents and emails relating to prescription (Rx) drug rebate negotiations between the Big 3 PBMs -- CVS Caremark, Aetna Express Scripts, and United Healthcare OptumRx -- and the Big 3 insulin drug manufacturers -- Sanofi, Novo Nordisk, and Eli Lilly.[5]

Three key features of signed rebate contracts in the appendices of the GW Report caught our eye:

1. PBMs offered a combinatorial bid menu featuring both exclusive and shared positions.
2. The standard basis for rebate offers was expressed as a % off unit list prices as measured by the publicly available wholesale acquisition cost (WAC).
3. After 2011, PBMs added to the bid menu an incremental rebate option for outright exclusion of named competing drugs.

Applying the descriptions in the GW Report to a taxonomy of market designs, we have conceptualized this exchange as a common value combinatorial auction.[6] It is a common value auction because Pharma's willingness to pay for formulary positions is profitability, a common value that both sides of this exchange can estimate.

---

5 United States Senate Finance Committee Staff Report, Charles E. Grassley, Chairman & Ron Wyden, Ranking Member, Insulin: Examining the Factors Driving the Rising Cost of a Century Old Drug Staff Report, (January 22, 2019), https://www.finance.senate.gov/imo/media/doc/Grassley-Wyden%20Insulin%20Report%20(FINAL%201).pdf

6 Nikhil Agarwal & Eric Budish, Market Design, National Bureau of Economic Research (Working Paper Series, No. 29367 October 2021), p. 7-8, http://www.nber.org/papers/w29367.pdf.; Peter Cramton, et. al, COMBINATORIAL AUCTIONS, MIT Press, (2006), Introduction. https://cramton.umd.edu/ca-book/cramton-shoham-steinberg-combinatorial-auctions.pdf

It is doubtful that PBMs ever have labeled their annual negotiations with Pharma as an auction. Nevertheless, it is an exchange, or market, with a particular design featuring mutually agreed upon rules. It has been repeated yearly now for over 20 years. We think it is reasonable to view it as a "vernacular" auction design that works, just as scholars have found that a lot of vernacular buildings designed centuries ago are consistent with modern architectural design theory.

### III. The PBM Winner's Determination Equation

The allegation that PBMs make anticompetitive formulary assignments considers only a winner's determination equation for a single position in any formulary therapeutic class for a single year. Normally, the objective of an auctioneer is to maximize proceeds for its principal, or in this case minimize total drug benefit costs. If only exclusive positions were up for auction, then the assignment consistent with that objective would be to assign exclusive positions to the drug with the lowest net unit price after unit rebates, regardless of gross rebates or expected market share.

There are a number of procompetitive reasons why PBMs make assignments other than just on the basis of net unit prices. The explanation starts with good auction design which dictates a bid menu of both exclusive and shared positions. Next we present the auction design rationale for shared positions and nuances in the PBM winner's determination equation.

First, offering a shared position option is equivalent to "set aside" bid packages in government procurement auctions designed to nurture emerging growth companies and long-term competition. Like government procurement auctions, the PBM auction is repeated each year. The goal of PBMs and their plan sponsors is not just single year drug benefit cost minimization, but long-term minimization.

The second rationale is that PBMs recognize that therapeutic equivalents are not perfect substitutes. Offering shared positions in any given therapeutic class can be viewed as a recognition by PBMs of the need to satisfy physician and patient preferences for some choice over strict adherence to total benefit cost minimization. In the GW Report Appendix, CVS Caremark mentioned lack of substitutability as it relates to "patient disruption" as a basis for not favoring lower net priced entrants over incumbents." [7]

Based on auction design theory, there is a third reason. Offering a combinatorial bid menu improves bid elucidation as its captures the superadditive or subadditive value of combinations.[8] The textbook example is an auction that offers a bid menu of airplane tickets to Hawaii, a week's stay in Hawaii, and a third package that combines both in order to capture the superadditive value of airplane tickets + a hotel room for some bidders.

In the PBM case, the profitability of a shared formulary position is subadditive as the number of allowed assignments increases. The reason is that average unit profitability of a shared position decreases with the number of assignments due to loss of production scale economies and increasing marketing costs. Generally, Pharma's willingness to pay as expressed as a % off WAC is lower for shared positions than for an exclusive position. This negative relation between unit rebate bids and the number specified in a shared position bid package has been substantiated by the GW Report.[9]

There is a fourth complication that is a consequence of only requiring a unit bid basis for rebates, which is good auction design in terms of encouraging bidding. Unit bids versus full lump sum bids reduces risks of overestimation of profitability and the "winner curse" associated with common value auctions. Profitability is the product of unit margin dollars times expected quantity demand of any given assignment. Assigning positions only on the basis of net unit prices creates the possibility that favored positions go to drugs with relatively little demand.

---

7 GW Report, Appendix, CVS Health Care (CVS Caremark), p.4.
https://www.finance.senate.gov/imo/media/doc/_FINAL%20PDF%20-%20CVS%20Caremark_Redacted.
8 Cramton, et. al.,*supra* p.33.
9 GW Report, Appendix, OptumRx, p.335 for example.
https://www.finance.senate.gov/imo/media/doc/OptumRx_Redacted.pdf

Our insight into this nuance came from an awareness of Google's ad position auction experience. It was Google's chief economist Hal Varian who introduced Google to the economics of position auctions.[10] The winners' determination equation required both $ per click and an estimate of click-through rates. Assigning ad positions only on $ per click could result in too many top positions going to high unit bidders with ads having little appeal and low click-through rates.

The need by Google to go beyond simple $ per unit bids is similar to the need for PBMs to go beyond assigning formulary positions solely on the basis of net unit prices. Because the objective of a PBM is to minimize total benefit costs, this requires PBMs to assign formulary positions on the combined bases of net unit prices and estimates of expected demand.

We believe that PBMs are aware of this nuance. However, our reading of the 1,700 redacted pages in the appendices to the GW Report confirmed our assessment of this exchange as a "vernacular auction" where the designers show little evidence of the economics literature on auction design. Pharma and PBM descriptions and computations found in the GW Report appendices are exceedingly crude compared to what you might find developed by Google to assess performance of its ad position auction. To us this is shocking, given the annual $100+ Billions in rebates exchanged in the Big 3 PBM auctions.

---

10 Hal Varian, Position Auctions, INTERNATIONAL JOURNAL OF INDUSTRIAL ORGANIZATION 25 (2007) 1163-1178. https://people.ischool.berkeley.edu/~hal/Papers/2006/position.pdf
.

## IV.     Auction Design Changes 2012 - 2014

The purpose of this section is to set the stage for our combinatorial auction model with an incremental rebate option tied to outright exclusion of named competitors. This addition was one of several implemented by PBMs between 2012 - 2014. The additions included

1. Adding gross rebates as a basis in the winners' determination equation.
2. Adding an incremental exclusionary rebate bid option tied to named drug competitor(s).
3. Adding a formulary administrative fee bid option expressed as a % off WAC.
4. Adding a price protection rebate bid option.

The Big 3 PBMs have self-reported a steady 8% average gross profit margin between 2017 - 2022. It was disclosed by hired consultants in a recent 126-page rebuttal to an FTC administrative complaint of unfair competition.[11]   That steady trend masks a potentially disruptive 2011 "patent-cliff", which PBMs offset via the above changes to their auction design.

PBMs started out as computer networking specialists who automated prescription claims processing by connecting retail pharmacy point of sales terminals to back-office health insurance mainframes. Sometime in the 1990s, PBMs added a look-up table to the point-of-sale software to automate switching off-patent brands to lower cost generics.

---
11 Carleton Report, (October 2024), p.75
https://compass-lexecon.files.svdcdn.com/production/files/documents/PBMs-and-Prescription-Drug-Distribution-An-Economic-Consideration-of-Criticisms-Levied-Against-Pharmacy-Benefit-Managers.pdf?dm=1728503869

Starting around 2000, the most popular therapeutic classes of drugs -- proton pump inhibitors, COX-2 inhibitors, second generation antihistamines, and statins -- started to see the entry of therapeutic equivalents with small differences in molecular structure. The opportunity for capturing some of the excess profits generated by patent-protected drugs was an order of magnitude greater than the surplus captured from switching off-patent brands to generics. Give PBMs credit for realizing that formularies could be used to create competition among drugs that otherwise were patent protected monopolists.

Our contribution to the PBM debate started in 2003, when we first disaggregated Express Scripts' 10-Q financials showing retained rebates exceeding 30%.[12] Even with the rebate retention rate leveling out at 10% over the next decade, PBMs were able to grow retained rebate dollars due to the increasing competition for formulary positions by patent protected but therapeutic equivalent drugs.

This competitive market for formulary positions is reflected in a doubling of PBM stock prices between 2006 to 2011 with the exception of 2008 due to the subprime mortgage financial crisis. Below is a graph of the split-adjusted stock price of the Express Scripts, the one pure play PBM over the period 2000 - 2018.[13]

PBMs had forecasting models predicting material loss of retained rebates dollars after 2011 due to a "patent-cliff" where a number of off-patent drugs would be forced to reduce list prices to levels of new generics competition. We believe that the impact of this new generic competition was reflected in the 2011 dip in Express Scripts' stock.

---

12 Lawrence W. Abrams, *Estimating the Rebate Retention Rate of Pharmacy Benefit Managers*, (Working Paper, April, 2003)
https://c1c0481a-8d34-49dc-ad66-d6c488c905a9.usrfiles.com/ugd/c1c048_a05060d8eead4427bc95ab12e9815e83.pdf
*Three Phases of the Pharmacy Benefit Manager Business Model*", (Working Paper, September 2017)
https://c1c0481a-8d34-49dc-ad66-d6c488c905a9.usrfiles.com/ugd/c1c048_e24040fcf1774f71b4438018d1f6d29f.pdf
13 Express Scripts Holding Co. (ESRX) Yearly Returns, https://www.1stock1.com/1stock1_264.htm

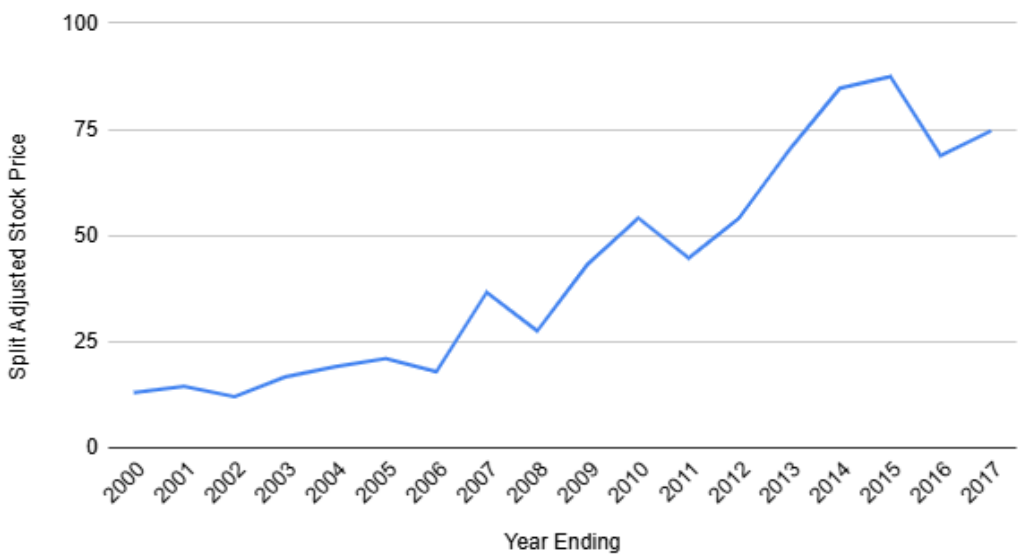

It was a clear warning sign to the Big 3 PBM CEOs that a new source of gross profits was needed. The problem was that small molecule drugs losing patent protection were ten times the Rx volume of new "rebatable" biologics. Plus, the PBMs were conceding to demands from the plan sponsors to lower rebate retention rates.

We have argued elsewhere that the dual threat of a small molecule patent cliff plus declining rebate retention rates motivated PBMs to collude to add gross rebates as a basis to the winners' determination equation.[14] We want to be clear. The allegation is that PBMs added gross rebates as a basis, not substituted gross rebates for net price after rebates. Most of the time, net prices outweigh gross rebates in the winners' determination equation.

---

14 Lawrence W. Abrams, *A Discovery Plan for Pharmacy Benefit Managers Collusion*, (Working Paper, October 2024) https://www.nu-retail.com/_files/ugd/c1c048_c0972a463aaf4d8196a1bfedb390fc51.pdf

Based on a graph of the so-called gross-to-net bubble in Rx drug prices, we pinpoint 2012 as the year for explicit communication among the Big 3 PBM CEOs to add gross rebates as a basis with an effective date of 2013.[15]

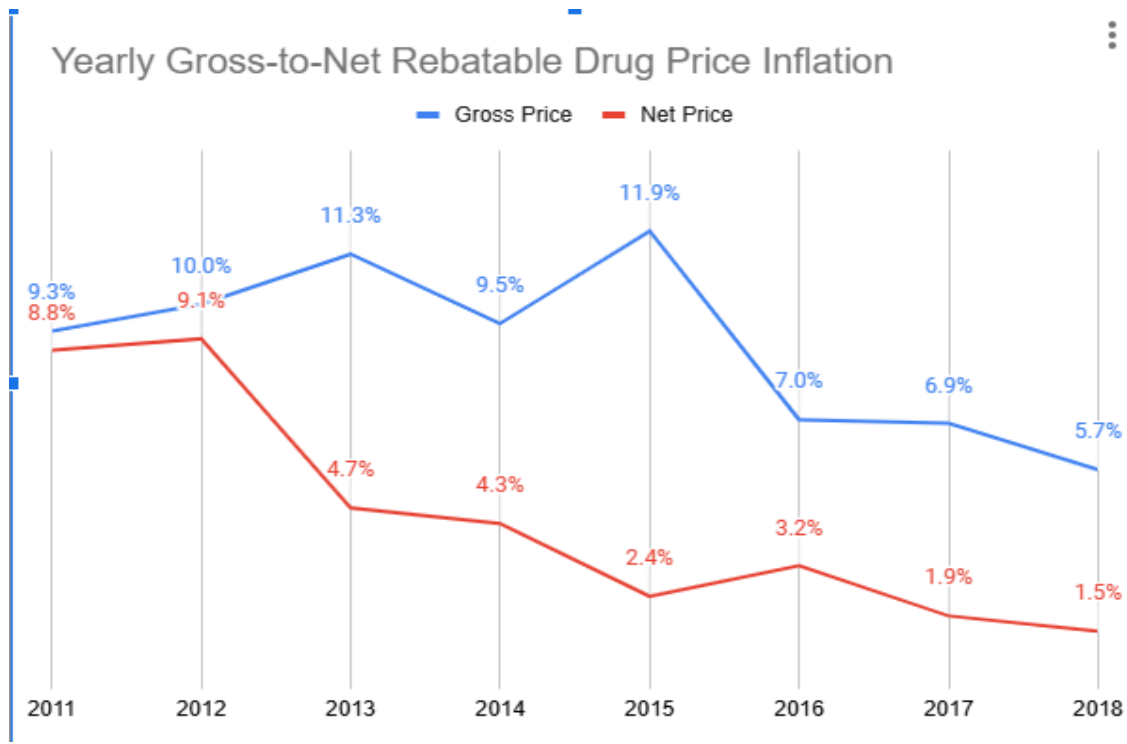

.Data from Adam Fein, *Drug Channel Blog, infra.*

Based on the GW Report, we believe PBMs added an incremental exclusionary rebate bid option for named drug competitor(s) between 2012 - 2014. One motive was the 2011 patent cliff. PBMs may have threatened exclusion, but in the end, would back off. At worst, a Pharma's drug would be assigned to highest copayment Tier 3.

___

15 Adam Fein, *Drug Channels Blog*, (June 14, 2017).
https://www.drugchannels.net/2017/06/new-data-show-gross-to-net-rebate.html
 Adam Fein, *Drug Channels Blog,* (January 29, 2019).
https://www.drugchannels.net/2019/01/drug-prices-are-not-skyrocketingtheyre
.

However, the PBM threat of a Tier 3 assignment became greatly diminished due to growing copayment assistance cards offered by Pharma after 2008. The addition of an incremental rebate for outright exclusion of names competitors can be seen as a response to copayment cards. In a 2013 Forbes article, an Express Scripts executive was quoted as saying as much.[16]

Below is a timeline of outright formulary exclusions showing that it was initiated by Caremark CVS in 2012 followed by Cigna - Express Scripts in 2014 and UnitedHealth Group - OptumRx in 2016.[17]

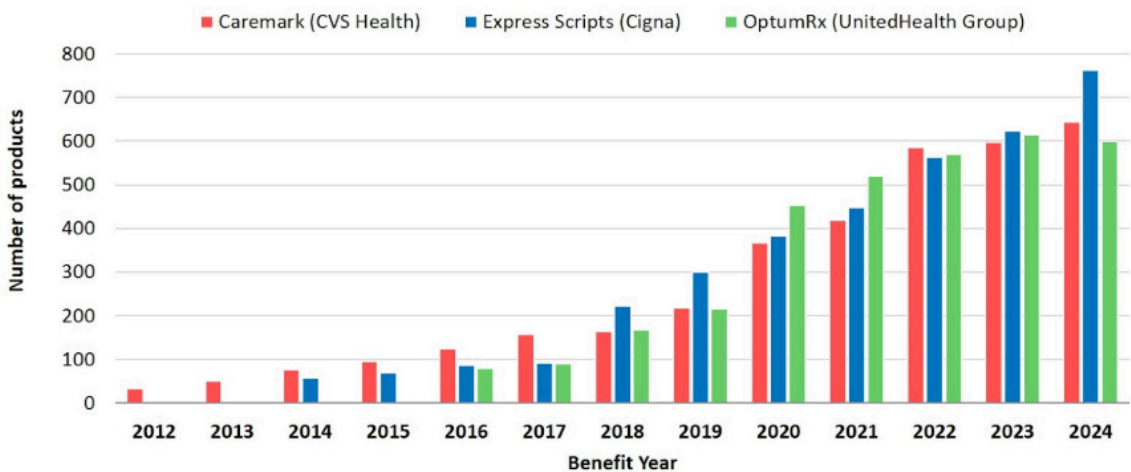

---

16 Ed Silverman, Bye, Bye Copay Cards? Why Express Scripts Is Excluding Dozens Of Drugs. FORBES, (October 12, 2013).
https://www.forbes.com/sites/edsilverman/2013/10/21/bye-bye-co-pay-cards-why-express-scripts-is-excluding-dozens-of-drugs/
17 Adam Fein, *Drug Channels Blog,* (January 9, 2024)
.https://www.drugchannels.net/2024/01/the-big-three-pbms-2024-formulary.html

To demonstrate that bid menus presented in our model have some connection to the real world, we present below a rare glimpse of actual PBM rebate bid menus from the appendices in the GW Report. First is a 2013 CVS rebate contract bid menu with Sanofi for its long-lasting basal Insulin drug Lantus.[18] Notice the low bids, which is what we would expect for a drug with no therapeutic equivalents at the time.

Effective 01/01/2014:

| Plan Type | 1 of 1 Manufacturer Status | 1 of 2 Manufacturer Status | Listed Formulary Status |
|---|---|---|---|
| Managed Plans - 2T | 2% | 2% | 2% |
| Managed Plans -3T | 3% | 3% | 3% |
| Highly Managed Plans | 4% | 4% | 4% |
| Closed Plans | 4% | 4% | 4% |

REBATES FOR LANTUS®
(INCLUDES ALL NDCs, STRENGTHS & PACKAGE SIZES)

Effective 01/01/2014:

| Plan Type | 1 of 1 Manufacturer Status | 1 of 2 Manufacturer Status | Listed Formulary Status |
|---|---|---|---|
| Managed Plans - 2T | 2% | 2% | 2% |
| Managed Plans -3T | 4% | 4% | 4% |
| Highly Managed Plans | 6% | 6% | 6% |
| Closed Plans | 7% | 7% | 7% |

REBATES FOR LANTUS® SoloStar®
(INCLUDES ALL NDCs, STRENGTHS & PACKAGE SIZES)

Contrast the above with a 2019 - 2022 OptumRx rebate contract bid menu with Sanofi for the same drug Lantus where now Lantus is facing competition from therapeutic equivalents from Novo Nordisk and Eli Lilly.[19] Notice now the "bid-down" between an exclusive position (1 of 1) and a shared position (1 of 2). Also notice the increase in combinatorial bid options due to the addition of administrative fees and cumulative WAC price protection rebates.

---

18  GW Report, Appendix, CVS Health Care, *supra*, p.55.
19  GW Report, Appendix, OptumRx, *supra,* p. 335.

## 5.1 PREFERRED
### 5.1.1 Lantus: (Effective 1/1/2019 through 12/31/2022)

| Benefit Design | Formulary Status | Manufacturer Drug Name: Lantus* | | |
|---|---|---|---|---|
| | | Highly Managed | Managed | Covered |
| Base Rebate Rate % | 1 of 1 manufacturer with Preferred Drugs | 75% | 65% | 50% |
| Base Rebate Rate % | 1 of 2 manufacturers with Preferred Drugs | 65% | 50% | 40% |
| Base Rebate Rate % | 1 of 3 manufacturers with Preferred Drugs | n/a | n/a | 26% |
| Administrative Fee | | 4.75% | 4.75% | 4.75% |
| Price Protection factor | | 4% | 4% | 4% |
| Baseline WAC Date | | 4/1/18 | 4/1/18 | 4/1/18 |
| Price Protection Year Start Date | | 1/1/19 | 1/1/19 | 1/1/19 |

Next we present a 2015 rebate contract bid menu between CVS and Novo Nordisk for its portfolio of insulin drugs.[20] We consider the bid-down from a 57.5% off WAC for an exclusive position to a 18% off WAC for a shared position to be a gross overestimate of the actual subadditive value of going from an exclusive to a shared position assignment.

We will argue in the next section that large bid-downs can be an anticompetitive strategy to exclude new entrants with low expected market share. We suggest how PBMs can prevent this anti-competitive strategy via incumbent bid-down limits.

---

20 GW Report, Appendix, Novo Nordisk, p.75.
https://www.finance.senate.gov/imo/media/doc/Novo_Redacted.pdf

**Exhibit A**
**Products, Rebates & Administrative Fees**
**(Percentage Rebates)**

The following Rebates and Administrative Fees shall be payable on Product dispensed to Participants by Participating Pharmacies:

| Product Name | NDC# | Strength | Package Size | Rebate Based on Formulary Status | | | | Administrative Fee |
|---|---|---|---|---|---|---|---|---|
| | | | | EGWP | Listed | 1 of 2 | Exclusive | |
| Novolin® | 00169-1833-11; 00169-1834-11; 00169-1837-11 | All Strengths | 10mL | 15% | N/A | 18% | 57.5% | 3% |
| NovoLog® | All NDCs | All Strengths | All Package Sizes | 15% | N/A | 18% | 57.5% | 3% |
| NovoLog® Mix 70/30 | All NDCs | All Strengths | All Package Sizes | 15% | N/A | 18% | 57.5% | 3% |

Finally, we present two examples of incremental bid options for outright exclusion of named competitors. Both are from a 2018 - 2020 contract between CVS Caremark and Sanofi. The first is a 15% incremental rebate bid option for outright exclusion of named competitors of Sanofi's Apidra, a rapid acting glulisine insulin.[21] The likely named competitors are Lilly's and Novo Nordisk's rapid acting insulins. The second is for the drug Lantus.[22] Note the low 2% to 3% incremental exclusionary rebate offer. Based on our calculations in the next section, we believe that the 2% to 3% incremental bid is a gross underestimate of the liquidated damages if that option bid is not accepted.

**EXHIBIT C-6**
**REBATES & ADMINISTRATIVE FEES**
Effective August 1, 2018 through December 31, 2020 (unless otherwise specified below)

A. **Base Rebates – Commercial Plans.** The following Base Rebate Percentages shall be applicable for Products dispensed by a Participating Pharmacy to Members of Commercial Plans:

A-1    Apidra®/Apidra SoloSTAR:

| BASE REBATES FOR APIDRA®/APIDRA® SoloSTAR® (INCLUDES ALL NDCs, STRENGTHS & PACKAGE SIZES) | | | | |
|---|---|---|---|---|
| Plan Type | 1 of 1 Manufacturer Status | 1 of 2 Manufacturer Status | Listed Formulary Status | Third Tier Status |
| Managed Plans - 2T | 66.0% | 41.0% | 36.0% | N/A |
| Managed Plans - 3T | 66.0% | 41.0% | 36.0% | 31.0% |
| Highly Managed Plans | 66.0% | 41.0% | 38.0% | 31.0% |
| Closed Plans* | 66.0% | 41.0% | 41.0% | 31.0% |
| Incremental Base Rebate For Additional Controls: (Not Applicable to Third Tier Status Rebates) | | | | |
| One Manufacturer of Competitive Products Excluded | | | 15.0% | |
| Two Manufacturers of Competitive Products Excluded | | | 15.0% | |

---
21 GW Report, Appendix, CVS Health Care, *supra,* p 72.
22 *id*. p 73.

A-3    Lantus and Lantus SoloSTAR:

| Plan Type | REBATES FOR LANTUS® and LANTUS SoloSTAR® (INCLUDES ALL NDCs, STRENGTHS & PACKAGE SIZES) | | | |
|---|---|---|---|---|
| | 1 of 1 Manufacturer Status** | 1 of 2 Manufacturer Status** | 1 of 3 Manufacturer Status ** | Third Tier Status |
| Managed Plans - 2T | 56.0% | 54.0% | 51.0% | N/A |
| Managed Plans -3T | 56.0% | 54.0% | 51.0% | N/A |
| Highly Managed Plans | 56.0% | 54.0% | 51.0% | N/A |
| Closed Plans* | 56.0% | 54.0% | 51.0% | N/A |
| Incremental Base Rebate For Additional Controls: (Not Applicable to Third Tier Status Rebates) | | | | |
| *One Manufacturer of Competitive Products is Excluded* | | | | 2.0% |
| *Two Manufacturers of Competitive Products are Excluded and applies only for 1 of 1 Manufacturer Status.* | | | | 3.0% |

*Managed Plans-2T, Managed Plans-3T, and Highly Managed Plans that adopt the Advanced Control Formulary™ will also be eligible for Closed Plan Rebate rates. Additionally, for clarity, Plans which do not otherwise qualify as Closed Plans, will receive Closed Plan Rebate rates with respect to the Lantus and Lantus SoloSTAR Competitive Category where Plan has opted into Exclusion Plan status for the Lantus/Lantus SoloSTAR Competitive Category.

## V.     An Combinatorial Auction Design for Formulary Positions

In this section, we develop a simple algebraic and graphic model of a combinatorial auction for formulary positions. The second version of the model includes a bid option for outright exclusion of named competing drugs. Based on Aghion and Bolton's (AB) foundational paper on vertical contracting with incomplete information, we view this bid option as liquidated damages (LD) -- Pharma's estimate of losses if a PBM decides not to accept, or breach, this incremental rebate offer.[23]

The market structure in the 1987 AB paper is remarkably similar to our model. The structure is a duopoly of an incumbent seller and an entrant (here there are two Pharma on the buy side) facing a single intermediary on the buy side (here it is a single PBM selling formulary positions). Without a PBM as an astute market designer, the market for formulary positions might be an AB case of contracts with incomplete information.

---

23 Aghion and Bolton, *supra*.

The AB paper proved that there are instances where an otherwise free market exchange might not maximize consumer welfare due the intermediary's failure to compare the incumbent's bid against an entrant's bid. The intermediary in the AB paper uncritically accepts the joint profit maximizing bid without soliciting a bid from a potential entrant with lower costs or willingness to accept lower margins. The AB paper gained prominence as a high quality economics critique of the Chicago School procompetitive bias toward exclusionary contracts between two parties.

Our approach inserts an auction designer into the AB case. The incremental exclusionary rebate bid is conceptualized as an AB liquidated damage (LD). The incumbent Pharma's exclusionary bid option is an estimate of its loss if its incremental exclusionary bid is not accepted. Unlike the AB case, there is a market designer whose function is to overcome incomplete information and seek out bids from potential entrants and supply additional estimates of shared position market shares required by the winners' determination equation.

We assume no principal-agent issues in our model due to the PBM misaligned reseller business model causing the PBM to assign on the basis of gross rebates rather than total benefit cost minimization. In reality, PBMs have a well-known agency issue stemming from their reseller business model which we have been at the forefront in exposing.[24]

We show that a rule of reason test still leaves open the possibility of an incumbent's untruthful bid-down strategy making it near impossible for an entrant to win a shared position assignment.[25]

---

24 Lawrence W. Abrams, *A Discovery Plan for Pharmacy Benefit Managers Collusion*, (Working Paper October 2024), https://www.nu-retail.com/_files/ugd/c1c048_c0972a463aaf4d8196a1bfedb390fc51.pdf
25 For an earlier treatment of this, see Lawrence. W. Abrams, *The Market Design for Formulary Position*, (Working Paper,March 2023),
https://c1c0481a-8d34-49dc-ad66-d6c488c905a9.usrfiles.com/ugd/c1c048_a0a9df7cf0de4a1094872ebd0f390941.pdf

To make this auction more incentive-compatible, we propose that the PBM limit incumbent bid-downs to the subadditive production and marketing costs of shared assignments. Without bid down limits, it is possible that the winners' determination equation will almost always be minimized by an exclusive assignment even if the entrant's rebate bid for a shared position causes its net price to be zero or negative.[26]

The model features a single PBM and two Pharma competing for formulary positions for their patented protected, but therapeutic equivalents. One is an incumbent with an expected dominant market share. The other is an entrant with the remaining share. The PBM conducts an auction using a combinatorial bid-menu allowing only unit bids expressed as a % off WAC. The PBM is tasked with making formulary assignments that minimize total expected benefit costs based on those bids and PBM's own estimates of expected demands for shared positions.

The first version of our model focuses only on two possible assignments: a Tier 2 exclusive position versus a Tier 2 shared position. If the entrant is not assigned a Tier 2 shared position, it is relegated to Tier 3 position with an immaterial market share. The possibility of a Tier 3 assignment is added in the next version of the model as the PBM can choose not to accept the incumbent's exclusionary LD option, but still accept its exclusive Tier 2 bid, thus relegating the entrant to Tier 3.

For simplicity, we assume WAC is the same for both. Because the total market in units T and WAC are on both sides of the winner's determination equation, we can drop them off and present the bid menu and winners' determination equation graphically below.

---

26 Abrams, *The Market Design for Formulary Position*, *supra*, p,19.

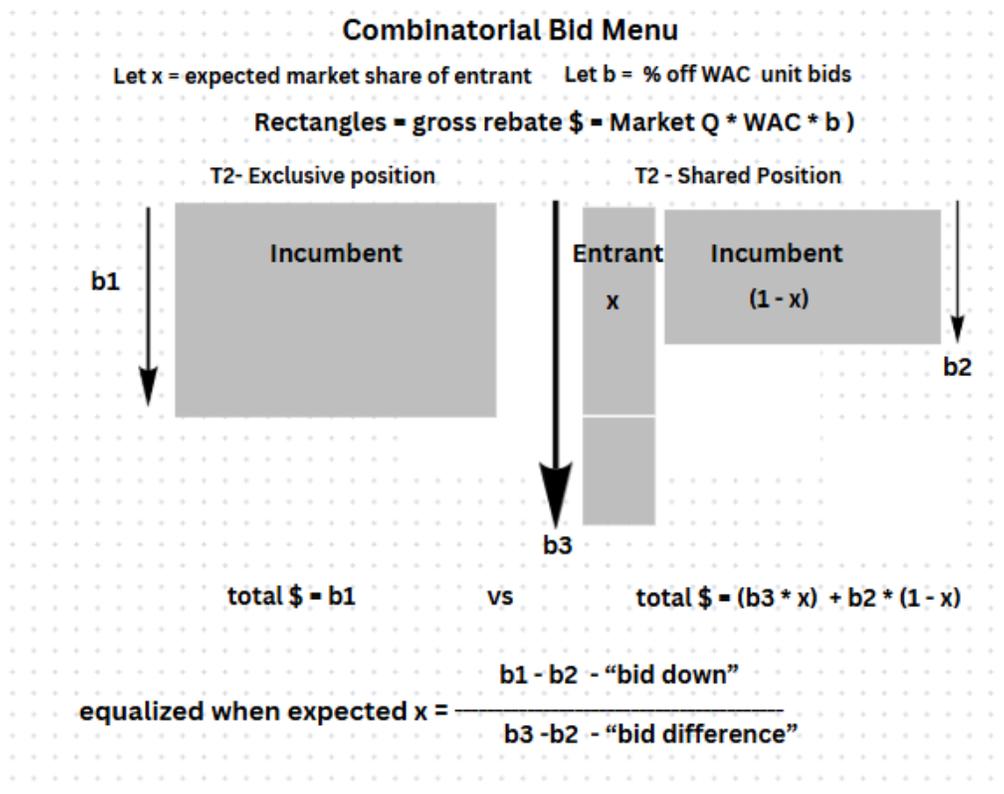

A PBM would favor an exclusive assignment over a shared assignment if

$$b1 > (b3 - b2) * x + b2$$

This equation highlights the possibility that even though the entrant's b3 bid for a shared position is higher than the incumbent's b1 for an exclusive position, its chances of winning a shared assignment are burdened by its low estimated market share and an incumbent's untruthful bid-down b2 for that shared position.

For example, let us go back to that 2015 contracted bid menu between CVS and Novo Nordisk for its portfolio of insulin drugs. The bid-down went from a 57.5% off WAC for an exclusive position to a 18% off WAC for a shared position.[27]  Consider now an entrant to this therapeutic class that both the PBM and the entrant agree does not have enough physician and patient familiarity to justify bidding on an exclusive position. Even if the entrant were to offer 90% off WAC, what would its expected market share have to be to warrant the PBM making a shared assignment?

|  |  |  |
|---|---|---|
| | (57.5% - 18%) | "bid down" |
| Equalizing market share x = 54.9% = | (90% - 18%) | "bid difference" |

Exhibit A
**Products, Rebates & Administrative Fees**
**(Percentage Rebates)**

The following Rebates and Administrative Fees shall be payable on Product dispensed to Participants by Participating Pharmacies:

| Product Name | NDC# | Strength | Package Size | Rebate Based on Formulary Status | | | | Administrative Fee |
|---|---|---|---|---|---|---|---|---|
| | | | | EGWP | Listed | 1 of 2 | Exclusive | |
| Novolin® | 00169-1833-11; 00169-1834-11; 00169-1837-11 | All Strengths | 10mL | 15% | N/A | 18% | 57.5% | 3% |
| NovoLog® | All NDCs | All Strengths | All Package Sizes | 15% | N/A | 18% | 57.5% | 3% |
| NovoLog® Mix 70/30 | All NDCs | All Strengths | All Package Sizes | 15% | N/A | 18% | 57.5% | 3% |

To make this auction more incentive compatible,  PBMs need to incentivize incumbents to bid to win, not bid-down so that entrants will lose. This can be achieved by limiting bid-downs as a % off WAC to the loss of production scale and the gain of marketing expenses due to sharing a formulary position as opposed to an exclusive position.

Because this is a common value auction based on profitability and Pharma's financials are publicly available, PBMs can estimate bid-down limits. Due to lawsuits and pressure from investors, Pharma now supplies gross to net revenue figures broken down by source -- formulary rebates, supply chain discounts, 340B, Medicaid, etc.. For fiscal 2023, Novo Nordisk reported a 34.2% average rebate gross to net.[28] This allows us to translate reported net revenue margins as gross revenue margins suitable for our auction model where rebate bids are a % off gross prices as measured by WAC.

---

27 GW Report, Appendix, Novo Nordisk, *supra,* p 75.

**Novo Nordisk - 2023 Margins on Gross and Net Basis**

|  | Margins on Net Basis | Margins on Gross Basis | Actual GtN Rebates | Estimate of Diseconomies of Gross Due to Shared Position |
|---|---|---|---|---|
| Gross Sales |  | 100% |  |  |
| Net Sales | 100.0% | 65.8% | 34.2% |  |
| Cost of Sales | 15.4% | 10.1% | >>> | 3% |
| Marketing | 24.4% | 16.1% | >>> | 7% |
|  |  |  |  | 10% |
| Contribution Margin | 60.2% | 39.6% |  |  |
| Max unit rebate | 39.8% | **60.4%** |  |  |

Source: Novo Nordisk 2023 Annual Report, p. 50, 55

Novo Nordisk, and brand Rx drug manufacturers in general, have relatively high R&D percentage costs and relatively low costs of goods sold with little production economies of scale. Indeed, the entry of therapeutically equivalent competition affects variable "physician detailing" and advertising costs more than average costs of goods sold.

So, the reported cost of goods sold and marketing cost of 15.4% and 24.4% of net sales translates to a 10.1% and 16.1% of gross sales, respectively. Based on these numbers, we build into our model a 2% production margin loss + 8% marketing margin loss = 10% total margin loss due to a shared position assignment. This translates into a 10% bid-down limit on an incumbent.

---

28 Novo Nordisk Annual Report 2023
https://www.novonordisk.com/content/dam/nncorp/global/en/investors/irmaterial/annual_report/2024/novo-nordisk-annual-report-2023.pdf

What an explicit LD does is change the winner's determination equation from a comparison of total rebates between the two possible position assignments to a comparison of incremental costs and benefits of position assignment switching. Limits on incumbent bid-downs make the design more incentive compatible in that both bidders are now incentivized to win and maximize their own switching costs.

**Combinatorial Bid Menu with incumbent bid down limited to subadditive value of shared position**
replaced by exclusionary liquidated damage bid limited to breach (switch to shared assignment)

|  |  | Combinatorial Bid Menu Rebate as % off WAC | |
|---|---|---|---|
|  |  | Exclusive T2 | Shared T2 |
| WAC $/Unit |  |  |  |
| WAC | Incumbent | b1 | 90%* b1 |
| WAC | Entrant | no bid | b3 |
|  |  |  |  |
| liquidated damages | Incumbent | b2 | n.a. |

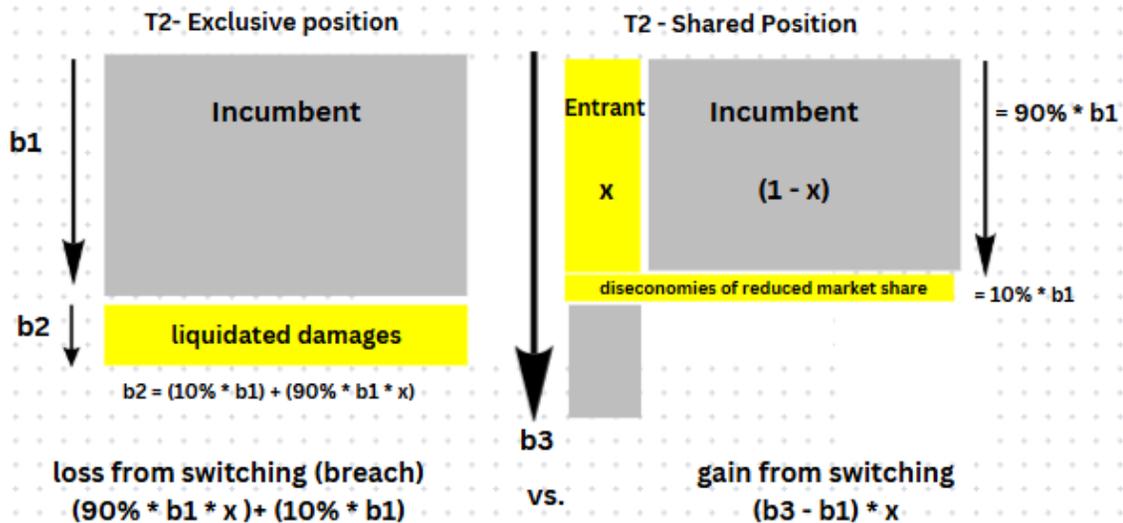

**Combinatorial Bid Menu**

Let x = expected market share of entrant    Let b = % off WAC unit bids
Rectangles = gross rebate $ = Market Q * WAC * b (yellow is rent-shifting)

T2- Exclusive position                    T2 - Shared Position

b1    Incumbent                           Entrant  Incumbent     = 90% * b1
                                            x      (1 - x)

                                          diseconomies of reduced market share  = 10% * b1
b2    liquidated damages
b2 = (10% * b1) + (90% * b1 * x)
                                          b3
loss from switching (breach)              vs.    gain from switching
(90% * b1 * x) + (10% * b1)                      (b3 - b1) * x

                          (10% * b1) - "bid down"
equalized when expected x = ─────────────────────────────
                          (b3 - 90% * b1) - "bid difference"

In reality, acceptance of the LD option also involves loss of rebates from a Tier 3 assignment. We estimate that a T3 position could garner as much as 10% of the total market of a therapeutic class. To cover that loss, the LD at a minimum should be (90% * b1 * 10%).  With a b1 in the range of 50% to 60% , the LD minimum bid should be around 4.5% to 5.4% off WAC. PBMs should reject any incremental exclusionary bid option less than that.

We conclude this section with a brief switch back to economist as analyst. Earlier, we characterized these negotiations as a "vernacular" auction, well designed in key areas based on the economic theories of good auction design. Yet, compared to the economics and computer science R&D of market designs of online companies like Google or Uber, we see little R&D invested by PBMs in their formulary position auction. To us, this is shocking, given the annual $100+ Billions in rebates exchanged in the Big 3 PBM auctions.

We base this only on a rare publicly available view of rebate negotiations presented in the heavily redacted GW Report. The U.S.Senate investigators' discovery plan itself was based on the question of who caused list price inflation, not how PBMs made formulary assignments. Our reading found no evidence of awareness of the bid-down problem. We found no evidence of any algebraic calculations or spreadsheets embodying nuances in the winners' determination equation discussed above.

The growth of outright formulary exclusions is correlated with the introduction of the LD option to the bid menu. As a result, we have no doubt that this addition helped hold down total benefit costs. Yet, an auction design with bid-down limits and algebraic evaluations of exclusionary LD bids might have reduced total benefit costs even further with less outright exclusions and more Tier 2 shared assignments.

**V. Bundled Rebates**

We conclude this paper with a market design approach to antitrust questions relating to bundled rebates. The practice involves Pharma tying an incremental % off a WAC bid for established drugs in its portfolio to a favored formulary assignment for an entrant drug in its portfolio. Pharma refers to this practice as "portfolio contracting" to minimize its association with an extensive history of antitrust cases involving exclusionary tying arrangements.

The standard antitrust approach to bundled rebates is to view it as a dispute between two Pharma. The formulary manager is passive. The winner's determination equation is assumed simply to be the lowest net unit price. The usual rule of reason test has been the price - cost test. Exclusionary conduct is deemed anticompetitive when the estimated net price after rebates of the tied drug is less than its estimated average cost of sale, known as predatory pricing.

As Salop has observed, it is generally hard to win antitrust tying arrangement cases based on this test.[29] It is "damn if you do, damn if you don't." A plaintiff's case can be dismissed on grounds that low prices improve consumer welfare. The case also can be dismissed on the grounds that pricing was not predatory and therefore the plaintiff's exclusion was due to something else.

In a market design approach, anticompetitive exclusionary conduct involves investigating both the bidders and the designer. We present the case below that non-standard bundled rebates are presumptive anticompetitive.

---

29 Stephen Salop "The Raising Rivals' Cost Foreclosure Paradigm, Conditional Pricing Practices, and the Flawed Incremental Price-Cost Test, 81 Antitrust L.J. 371-421 (2017) https://scholarship.law.georgetown.edu/cgi/viewcontent.cgi?article=2632&context=facpub#:~:text=Myriad%20types%20of%20business%20conduct,may%20not%20be%20so%20obvious

A traditional antitrust approach only looks at what is. An auction design approach directly addresses the "but-for" question with quantitative estimates of net unit prices and the winners' determination equation of a standard PBM auction design. An auction design approach would demand discovery of evidence that the plaintiff and the PBM made attempts to hold a standard auction limited to individual drug unit rebate bids.

There is a pending 2022 lawsuit by Regeneron alleging Amgen bundled its self-injectable PCSK9 inhibitor diabetes drug Repatha with two other established drugs in its portfolio - Otezla and Enbrel.[30] Regeneron chose to file their case under Sherman Act, Sect. 2 "attempt to monopolize." as opposed to Sherman Act, Sect. 1 "restraint of trade", the usual statute for cases involving tying or bundling arrangements. Regeneron alleged that Amgen's estimated $80 Million / year tied amount made it impossible for it to counter with a competitive bid for its own PCSK9 inhibitor diabetes drug Praulent without going below unit cost of sale.

Below we present plausible bid menus and associated estimated dollars of winning bids from a standard auction design. Our conclusion is that a standard auction could have exceeded the $80 Million / year bundled rebate. If Regeneron had evidence that it at least petitioned the PBM for a standard auction, it still might have a case. However, it would be one based on an unfair market design not exclusion due to predatory pricing as we believe that Regeneron could have won a shared position if it had bid about 70% off list under a PBM standard auction.

---

30 Big Molecule Watch, Regeneron vs Amgen,
https://www.bigmoleculewatch.com/wp-content/uploads/sites/2/2022/06/Regeneron-Antritrust-Complaint-vs-Amgen-May-27-2022-3.pdf
  Regeneron Pharmaceuticals, Inc. v. Amgen Inc., 22-cv-697
https://casetext.com/case/regeneron-pharm-v-amgen-in

| Amgen's Bundled Rebate Offer | | | | |
|---|---|---|---|---|
| Exclusive Tier 2 | | Repatha | Enbrel | Otezla |
| Rebate as % off WAC | | n.a. | 0 | 0 |
| | | Tying | Tied | Tied |
| Bundled Rebate Bid - $ | | $80 Million / year | | |
| "But if" Competition for Exclusive Position - Estimate | | | | |
| | | Amgen | Regeneron | |
| Exclusive Tier 2 | | Repatha | Praulent | |
| Rebate as % off WAC | | 65% | 70% | |
| Rebate $ Amount | | $96 Million / year | | |
| Based on | | | | |
| Estimated US Total Gross Dollar PCSK-9 Market - Mil | | | | $900 |
| PBM - ESI Commercial Formulary Market Share - % | | | | 15.27% |
| PBM - ESI Commercial Formulary Market Share - Mil | | | | $137 |
| Winning rebate - % off WAC | | | | 70% |
| Individual Drug Winning Bid - $ | | $96 Million / year | | |

The prescription drug industry is characterized by relatively high R&D costs coupled with low manufacturing costs. Costs of sale as a % of WAC today generally run around 10% to 15% with little production economies of scale. As we have discussed in the previous section, *bona fide* bid-downs for a shared formulary position are in the 10% range due mostly to increased marketing costs, not manufacturing costs. A well-designed formulary position auction should be able to get a bid in the range of 70% to 85% off WAC from an entrant for a shared position. This is not a material difference from what a PBM could get by accepting a non-standard secretive bundled rebate offer.

Officially, only the law firm representing CVS Caremark acknowledged in the GW Report that the practice exists among the Big 3 insulin manufacturers. The CVS Caremark law firm states that the practice was initiated by a Pharma but made no statement about acceptance by CVS Caremark.[31]

---

31 GW Report, Appendix, CVS Health Care, supra, p.14.

The other PBM lawyers only state that the "typical" basis for rebates is a single % off unit WAC. However, there are a number of emails in the GW Report between top Pharma executives referencing the practice of "bundling" and "portfolio contracting." The GW Report even let slip through a few explicit bundled rebate bid proposals. All in all, bundled rebate references in the GW Report have an aura of secrecy.

The most references in the GW Report appendices were Novo Nordisk tying incremental rebates for its established drugs Novolog, Novolin, and Victoza to favored assignment of its long-lasting basal insulin drug Toujeo. There was one reference to Sanofi tying incremental rebates for its dominant basal insulin drug Lantus to a favored formulary assignment for its entrant ephedrine pen Auvi-Q.[32]

The majority of references were made by Sanofi alluding to bundling practices by its archrival Novo Nordisk. Starting in 2016, Sanofi's long-acting basal insulin drug Lantus faced competition from Levemir, a new entrant from Novo Nordisk. Sanofi made six references alluding to its archrival Novo Nordisk tying incremental rebate bids for its established portfolio drugs Novolin, Novolog, and Victoza to favorable assignment for its entrant drug Levemir.[33] It is as if Sanofi purposely supplied the Senate investigators internal mails referencing this practice to "rat out" Novo Nordisk.

Sanofi itself makes two references to tying incremental rebates for its established drug Lantus to favorable assignments for its entrant drug Toujeo.[34] Here is a screen shot from the Sanofi appendix of a 2015 bid executive powerpoint presentation actually arguing against a bundled rebate offer on the grounds that that might trigger a higher Medicaid best price.[35]

---

32 GW Report, Appendix, Sanofi,, p 334
33 id., pp.112,114,121,129,218, 396.
34 GW Report, Appendix, Sanofi, *supra*., pp.313,380.
35 id., p. 235.

## Lantus Auvi-Q Bundle Discussion

**Background**
- Express Scripts and Envision have Contracts that increase Lantus rebates if Auvi-Q is added to formulary thus creating a bundled arrangement
- Current Terms: Ex. ESI Lantus 1 of 2 26%, 1 of 1 32% Auvi-Q 1 of 2 – 30%, 1 of 1 – 65%

**Situation/Risk**
- The potential of adding Toujeo will create a "triple" product bundle
- *Discounts are allocated based sales of product with the potential of the 65% dicount impacting the reported Governemnt best price on Toujeo and potential for increasing the Medicaid mandated rate*
- Government Pricing Compliance/Operation Risk in calculating bundle

**Discussion**
- Amend Contract to remove Auvi-Q from bundled arrangement

Rationale: Reduces Risk of Government impact and Compliance

Risk: Exposes Auvi-Q at ESI and Envision. ESI ~$55M Net, ~$1M Net

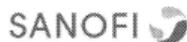
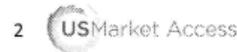

FOR INTERNAL USE ONLY: DO NOT DISTRIBUTE

While Sanofi's appendix has a number of references to Novo Nordisk's rebate bundling strategies, we could find no references in the Novo Nordisk appendix either to a competitor's use or their own. We found one reference in the Lilly appendix to a Novo Nordisk offer and two references to their own offers.[36]

It is not clear how prevalent non-linear formulary rebate bids are today. A Lina Khan led FTC might have just squashed the practice with a March 2024 lawsuit against Amgen to block its bid to acquire Horizon Therapeutics. Separately the FTC filed an administrative complaint of "unfair competition" under Section 5 of the FTC Act, 15 U.S.C. § 45.

___________________

36 GW Report, Appendix, Eli Lilly, pp 39, 49,105.
https://www.finance.senate.gov/imo/media/doc/Eli%20Lilly_Redacted%20v1.pd

The allegation was that Amgen intended to use bundling to win formulary inclusion for Horizon's up and coming drugs Tepezza and Krystexxa. Amgen signed a consent decree not to engage in bundling. The FTC then allowed the merger to proceed proclaiming that "this case was the FTC's first litigated challenge to a pharmaceutical merger in more than a decade."[37]

We conclude with this final thought: an effective way to squash the practice would be for Congress to consider adding a clause to pending PBM reform legislation banning bundled rebates or other non-linear bids by parties in Medicare Part D contracts.

---

37 Federal Trade Commission, "FTC Approves Final Order Settling Horizon Therapeutics Acquisition Challenge, December 14, 2023.
https://www.ftc.gov/news-events/news/press-releases/2023/12/ftc-approves-final-order-settling-horizon-therapeutics-acquisition-challenge